\documentclass[11pt,english,aps,journal=mamobx,manuscript=article,maxauthors=15,biblabel=plain]{revtex4-1}
\usepackage[T1]{fontenc}
\usepackage[latin9]{inputenc}
\usepackage[a4paper]{geometry}
\usepackage{amsmath}
\usepackage{amssymb}
\usepackage{graphicx}

\makeatletter
\usepackage{babel}

\makeatother

\usepackage{babel}
\begin{document}

\title{The Formation and Structure of Olympic Gels}

\author{J. Fischer}

\altaffiliation{Present address: Bio Systems Analysis Group, Jena Centre for Bioinformatics (JCB) and Department for Mathematics and Computer Sciences, Friedrich Schiller University of Jena, 07743 Jena, Germany.}

\affiliation{Leibniz Institut für Polymerforschung Dresden, Hohe Straße 6, 01069
Dresden, Germany.}

\author{M. Lang}

\email{lang@ipfdd.de}

\affiliation{Leibniz Institut für Polymerforschung Dresden, Hohe Straße 6, 01069
Dresden, Germany.}

\author{J.-U. Sommer}

\affiliation{Leibniz Institut für Polymerforschung Dresden, Hohe Straße 6, 01069
Dresden, Germany.}

\affiliation{Technische Universität Dresden, Institute of Theoretical Physics,
Zellescher Weg 17, D-01069 Dresden, Germany.}
\begin{abstract}
Different methods for creating Olympic gels are analyzed using computer
simulations. First ideal reference samples are obtained from freely
interpenetrating semi-dilute solutions and melts of cyclic polymers.
The distribution of pairwise concatenations per cyclic molecule is
given by a Poisson-distribution and can be used to describe the elastic
structure of the gels. Several batches of linear chains decorated
with different selectively binding groups at their ends are mixed
in the ``DNA Origami'' technique and network formation is realized.
While the formation of cyclic molecules follows mean field predictions
below overlap of the precursor molecules, an enhanced ring formation
above overlap is found that is not explained by mean field arguments.
The ``progressive construction'' method allows to create Olympic
gels with a single reaction step from a concentrated mixture of large
compressed rings with a low weight fraction short chains that are
below overlap concentration. This method, however, is limited by the
difficulty to obtain a sufficiently high degree of polymerization
of the large rings.
\end{abstract}
\maketitle

\section{\label{sec:Introduction}Introduction}

``Olympic gels'' (OGs) \cite{key-1} are networks made of cyclic
polymers (also called polymer rings) that are not linked by chemical
cross-links. Instead, the permanent entanglement between concatenated
rings establishes a three dimensional network structure similar to
the structure of the Olympic rings. Due to the lack of cross-links,
these samples were considered to be ideal model networks that might
allow for an unperturbed analysis of the effect of entanglements in
a polymer network \cite{key-2}. Recently \cite{key-3}, it was shown
by computer simulations that OGs show unusual swelling properties,
since networks of longer strands swell less than networks made of
short cyclic molecules at otherwise identical preparation conditions.
This was explained \cite{key-3} by the desinterspersion of overlapping
non-concatenated rings, that causes a large non-affine contributions
to the swelling of these gels at the available low average number
of concatenations per ring.

\begin{figure}
\begin{centering}
\includegraphics[width=1\columnwidth]{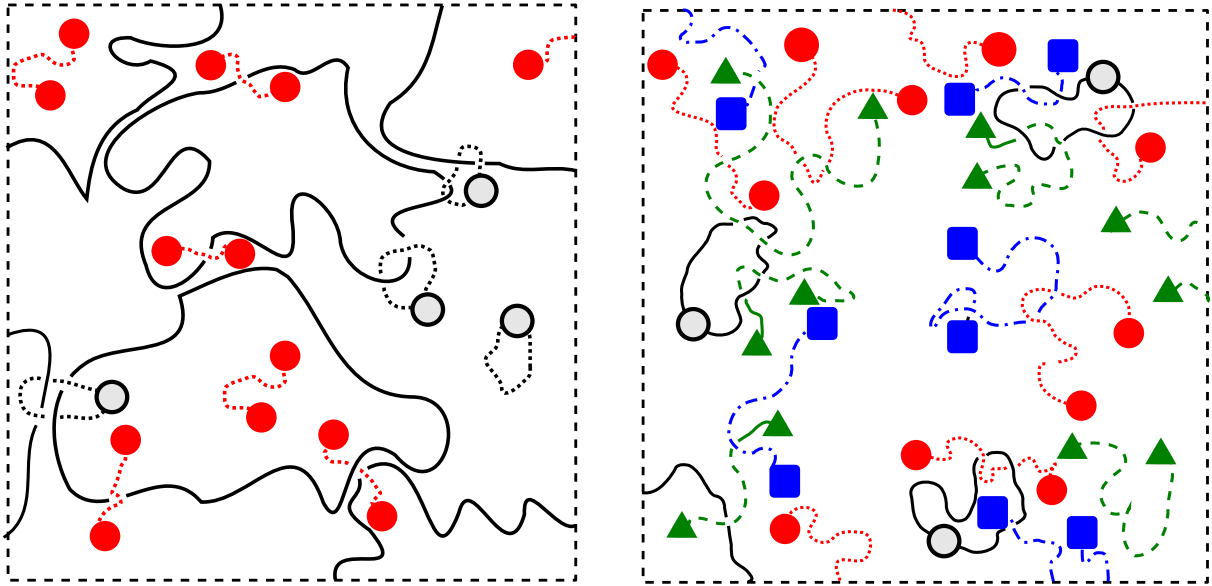}
\par\end{centering}

\caption{\label{fig:sketch_methods}Progressive construction (left) and DNA-Origami
(right). Lines represent polymers, dots of same color (color online
all Figures) are reactive end-groups of same type and gray dots with
black boundary are reacted polymer end-groups.}
\end{figure}

Even though these OGs are, therefore, a very interesting model system
to understand the physics of entanglements, these network have not
yet been synthesized. The reason for this can be understood from considering
the geometry of cyclization: Let us consider a mono-disperse solution
of linear $N$-mers that are long enough for cyclization. In order
to form an OG, the cyclic molecules have to be at sufficient overlap
such that a sufficiently large average number of concatenations per
ring, $f_{n}$, can be established. Let $P\approx\phi R^{3}/(N\mbox{v}_{0})$
denote the number of overlapping linear chains of the same degree
of polymerization $N$, monomeric unit volume $\mbox{v}_{0}$ and
size $R$ at a polymer volume fraction of $\phi$. Gel formation,
therefore, requires $P\gg1$ and a weight average of at least two
connections per molecule. However, the probability that the $N$-mer
forms a ring is only of order $\approx1/(2P)$, since there is only
one opposite end of the same chain to react with in the pervaded volume.
Therefore, even if all $P/(2P)$ overlapping rings were mutually concatenated,
these rings would not form a gel, since there is less than one connection
per ring. Interestingly, this discussion depends only on $P$ (which
even cancels out) and not explicitly on $\phi$ and $N$. Therefore,
one cannot tune $N$ or $\phi$ such that an OG might be obtained.
Instead, one has to modify the synthesis to allow for the formation
of OG's. In this respect, Raphael \emph{et al.} \cite{key-4} (based
upon an idea of De Gennes \cite{key-1}) and Pickett \cite{key-5}
proposed two different methods for preparing OGs as sketched in Figure
\ref{fig:sketch_methods}.

The first method by De Gennes \cite{key-1} is called ``progressive
construction'', since the network structure is built up in several
steps. First, large ring polymers are made at very dilute concentrations
in order to suppress the competing growth of linear chains. Next,
the solution is concentrated above the overlap concentration of the
rings. Then, end-functionalized linear chains are allowed to diffuse
into the solution, whereby the concentration of linear chains is chosen
below the overlap concentration of the linear chains in order to promote
cycle formation. When the mixing equilibrium is reached, the reaction
is started and a fraction of the surrounding cyclic polymer is entrapped
by the closure of the linear polymers, see left of Fig \ref{fig:sketch_methods},
which may lead to an OG, if a sufficiently large number of connections
between the long rings are established.

The ``DNA-Origami'' approach of Pickett \cite{key-5} requires the
fabrication of a large set of different selectively binding end-groups.
Pickett proposed to isolate a large set of different DNA fragments
and to attach the ends of one chain to the two strands of one of these
fragments. At solvent conditions under which DNA denaturates, the
small DNA fragments will open up and the polymers decorated with the
single strands of these fragments on both ends will form a linear
solution of polymers. If the solvent conditions are modified such
that the opposite DNA fragments stick together, the polymers can form
rings upon selectively binding with the corresponding DNA fragment.
If the number of different fragments is sufficiently large, OG's are
obtained because chains of same type of fragment will be below overlap
concentration and thus, ring formation will dominate the growth of
linear chains.

In addition to these methods, one could also use topoisomerase to
create OGs of nearly perfect structure from semi-dilute solutions
of cyclic DNA. We created OGs in similar manner by allowing the polymer
strands to interpenetrate freely in our computer simulations. The
data of these simulations is used below as ideal reference systems
due to the missing polydispersity of the cyclic molecules and the
absence of linear chains. In the present paper, we test these different
approaches concerning their applicability to produce OGs and concerning
the quality of the network structure that can be obtained in this
way both analytically and by computer simulations.

\section{\label{sec:Simulations}Computer Simulations and Analysis}

We use the bond-fluctuation model (BFM) \cite{key-6,key-7} to simulate
solutions of linear chains and rings. This method was chosen, since
it is is known to reproduce conformational properties and dynamics
of melts \cite{key-8,key-9} and semi-dilute solutions \cite{key-10,key-11}
and polymer networks \cite{key-12,key-13}. In this method, each monomer
is represented by a cube occupying eight lattice sites on a cubic
lattice. The bonds between monomers are restricted to a set of 108
bond vectors which ensure cut-avoidance of polymer strands by checking
for excluded volume. Monomer motion is modeled by random jumps to
one of the six nearest lattice positions. A move is accepted, if the
bonds connecting to the new position are still among the set of 108
bond vectors and if no monomers overlap. All samples of the present
study were created in simulation boxes with periodic boundary conditions.
A-thermal solvent is treated implicitly by empty lattice sites.

In our work, we discuss two different series of simulations in order
to analyze ideal OG and the competition between growth of linear chains
and ring formation (DNA-Origami). The results of these simulations
serve as input to analyze the ``progressive construction'' method.
The details of sample preparation of these simulation series are summarized
at the beginning of the corresponding sections.

For analysis of network connectivity, rings were first simplified
at conserved topology in one additional simulation run by removing
monomers $i$ from the chain contour, whenever monomers $i$, $i+1$
and $i-1$ formed a tight triangle through which no bond of any other
molecule could pass \cite{key-14}. Next, we determined a regular
projection of any pair of overlapping rings with minimum number of
intersections as also described in \cite{key-14}. Finally, we computed
the Gauss code of the projection as input for the Skein-Template algorithm
of Gouesbet \emph{et al.} \cite{key-15}. The types of the knots and
pairwise links formed were analyzed by using the resulting HOMFLY
polynomials \cite{key-16}. Data of previous work \cite{key-17} show
that the effect of Brunnian links or similar structures that are not
(entirely) detected by a pairwise linking analysis should be ignorable
for the degrees of polymerization of our study. Therefore, we restricted
our connectivity analysis to pairwise links only. The resulting connectivity
matrix between pairs of rings is used to analyze the network structure,
weight fraction of gels, the gel point position, and the amount of
the elastically active rings.

\section{Ideal Olympic gels\label{sec:Ideal-Olympic-gels}}

Interpenetrating solutions of mono-disperse cyclic polymers are considered
to serve as an ideal reference system to understand the formation
of OGs, since in these samples, concatenation is in equilibrium with
the polymer conformations in solution at the particular polymer concentration.
Our simulations of these ideal OGs cover an array of different degrees
of polymerization $N$ = 16, 32, 64, 128, 192, 256, 384, 512, 768,
and 1024 and polymer volume fractions of $\phi$ = 0.5, 0.375, 0.25,
0.1875, 0.125, 0.0625, and 0.03125 as described previously \cite{key-11}.
The numbers of rings per sample varied from 512 to 4096, resulting
in a simulation box size between $128^{3}$ and $512^{3}$ lattice
sites. Periodic boundary conditions were applied in all space directions.
During equilibration, additional ``diagonal'' moves were allowed
that do preserve connectivity of a ring but allow for a change in
the topology of overlapping rings, since all entanglements are switched
off. After equilibration, all ``x-traps'' (pairs of bonds that mutually
block each other's motion - these arise here from switching off diagonal
moves) \cite{key-18} were removed while returning to the original
set of moves and network connectivity is analyzed as described in
the previous section.

In our preceding work \cite{key-11}, we analyzed only the average
number of concatenated rings per ring polymer, the number average
``functionality'' of the rings, $f_{n}$, in a solution of mono-disperse
interpenetrating rings in order to develop a model for the conformations
of rings in melt or solution. It was found that 
\begin{equation}
f_{n}\approx\gamma\phi^{\nu/(3\nu-1)}N\left(1-P_{\mbox{OO}}\right),\label{eq:fn}
\end{equation}
with a numerical constant $\gamma\approx0.034\pm0.001$ and a cut-off
$\left(1-P_{\mbox{OO}}\right)$ for very small $\phi^{\nu/(3\nu-1)}N$
based upon the probability for non-concatenation $P_{\mbox{OO}}\approx\exp\left(-\left(\phi^{\nu/(3\nu-1)}N-a\right)/N_{\mbox{OO}}\right)$.
Here, $N_{\mbox{OO}}$ is the cross-over degree of polymerization
(as defined for extrapolating to $\phi=1$) that distinguishes between
the non-concatenated regime $N<N_{\mbox{OO}}$ and the concatenated
regime $N>N_{\mbox{OO}}.$ Similar to knotting, a second parameter
$a$ is used that can be understood as effective minimum degree of
polymerization to make concatenation possible (with respect to a particular
environment). However, for poly-disperse systems, gelation requires
\cite{key-19} a \emph{weight average} functionality $f_{w}$ of two
\begin{equation}
f_{w}=2\label{eq:pc}
\end{equation}
instead of $f_{n}=2$. Therefore, we have to determine $f_{w}$ and
the distribution of concatenations in order to estimate the position
of the gel point.

\begin{figure}
\begin{centering}
\includegraphics[angle=270,width=1\columnwidth]{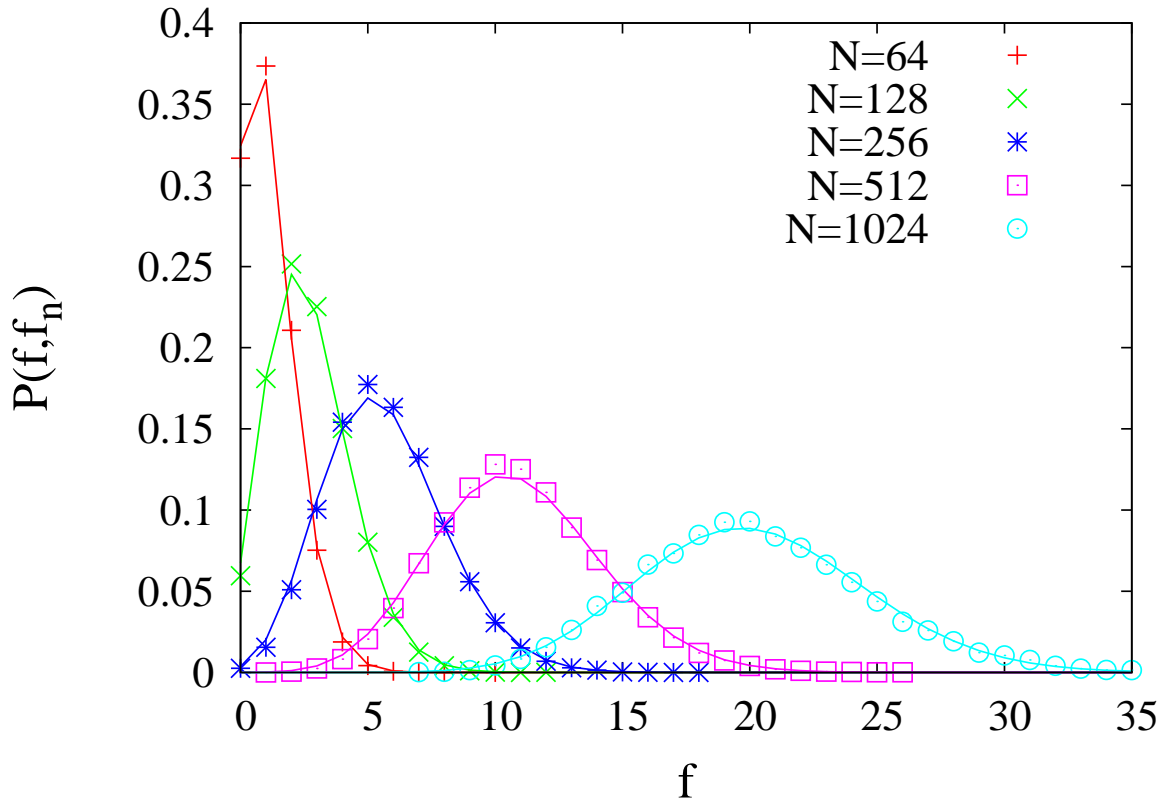}
\par\end{centering}

\caption{\label{fig:Distribution-of-the}Distribution of the number of concatenations
$f$ per ring (data points) in mono-disperse solutions of interpenetrating
rings at $\phi=0.5$. The lines are computed using equation (\ref{eq:poisson})
and the average number of concatentations $f_{n}$ as the only adjustable
parameter.}
\end{figure}

\begin{figure}
\begin{centering}
\includegraphics[angle=270,width=1\columnwidth]{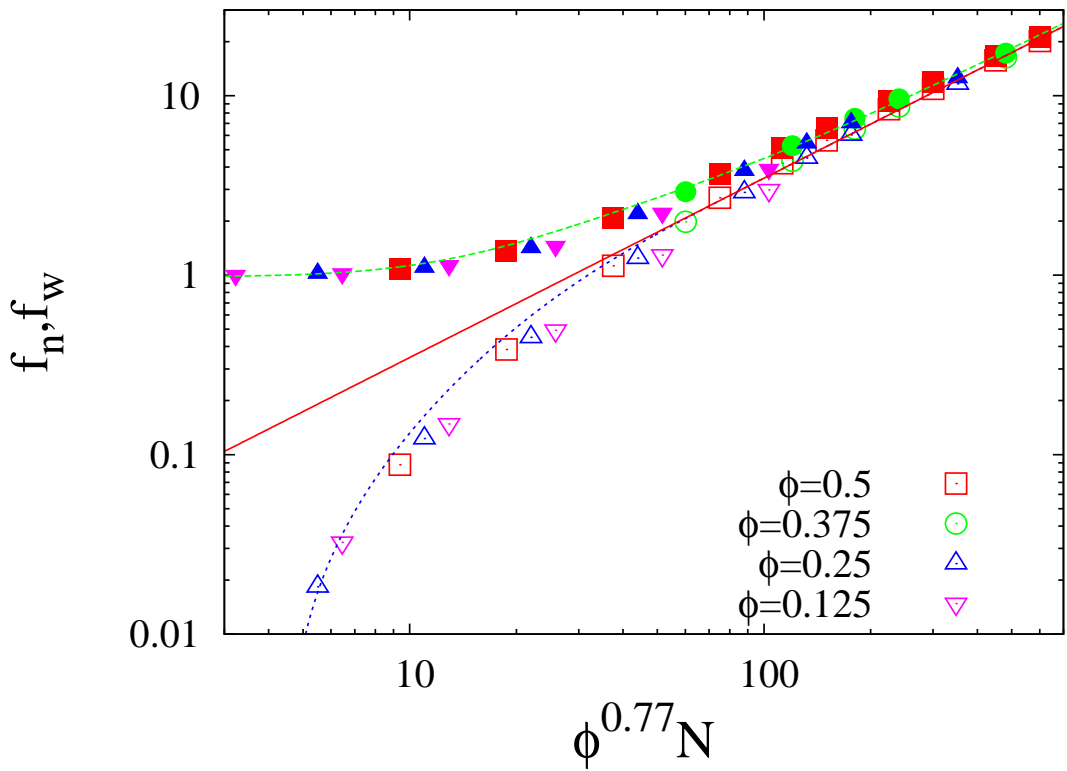}
\par\end{centering}

\caption{\label{fig:Number-average-}Number average $f_{n}$ (hollow symbols)
and weight average $f_{w}$ (filled symbols) number of concatenations
of cyclic polymers in mono-disperse solutions of cyclic polymers.
The continuous line is a linear increase, the dotted line is equation
(\ref{eq:fn}), and the dashed green line is equation (\ref{eq:fw}).}
\end{figure}

In a previous work \cite{key-20}, it was argued that the number of
concatenations per ring is related to the area of the minimal surface
bounded by a ring. A rather narrow distribution of the area of the
minimal surfaces of rings of same degree of polymerization at the
same preparation conditions can be assumed because of the dominance
of the boundary region of the area (see ref. \cite{key-20}). Let
us assume that concatenation is a random process that occurs with
equal probability per polymer strand that passes through this rather
constant area of the minimum surface. Under these conditions, the
distribution of concatenated states is expected \cite{key-32} to
be described by a Poisson distribution 
\begin{equation}
P(f,f_{n})=\frac{f_{n}^{f}}{f!}\,\mathrm{e}^{-f_{n}}\label{eq:poisson}
\end{equation}
around the number average number of concatenations, $f_{n}$, per
ring. Figure \ref{fig:Distribution-of-the} shows an excellent agreement
between equation (\ref{eq:poisson}) and the data of the computer
simulations independent of the degree of polymerization $N$ and the
average number of concatenations $f_{n}$. The weight average functionality
of the above Poisson distribution \cite{key-21} is 
\begin{equation}
f_{w}=f_{n}+1\label{eq:fw}
\end{equation}
and fits well to the simulation data as shown in Figure \ref{fig:Number-average-}.
Note that the cut-off for non-concatenation, $P_{OO}$, is only a
weak correction at the gel point (of order 10\%) an can be safely
ignored for well developed gels.

\begin{figure}
\begin{centering}
\includegraphics[angle=270,width=1\columnwidth]{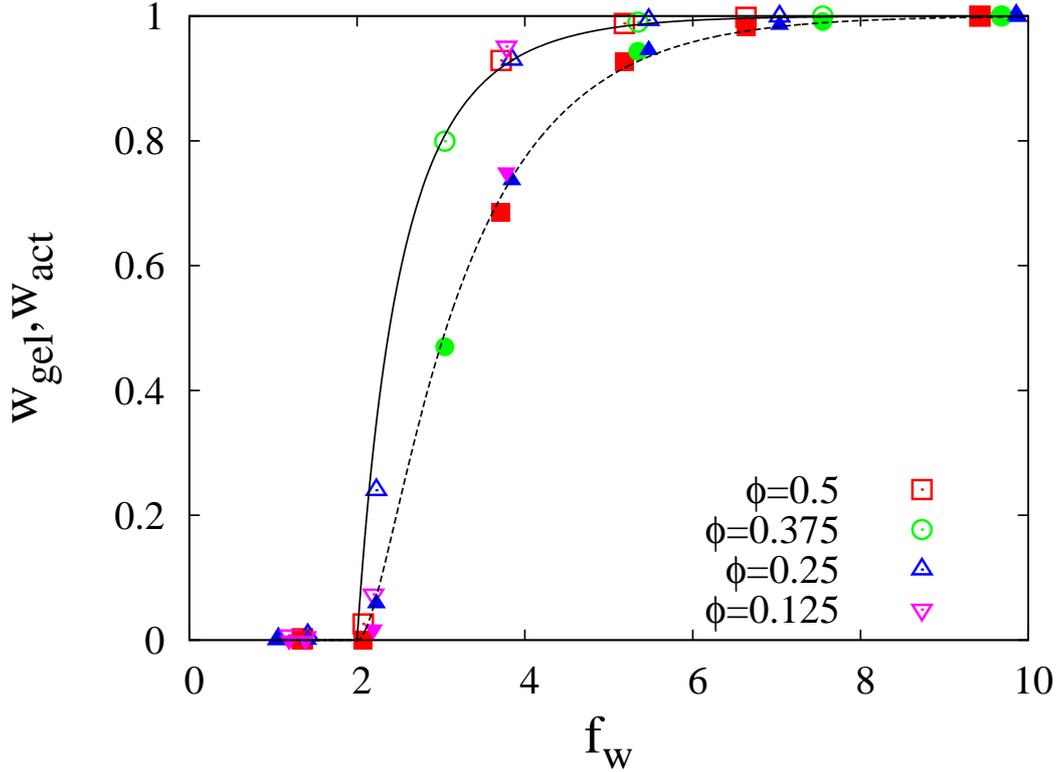}
\par\end{centering}

\caption{\label{fig:The-weight-fraction}The weight fractions of the largest
connected cluster of concatenated rings (hollow symbols) and the elastically
active material (full symbols) as function of the weight average functionality
of the rings. Continuous line is weight fraction of gel, equation
(\ref{eq:w_gel}), and dashed line is weight fraction of active material,
equation (\ref{eq:w_act}). }
\end{figure}

The above estimate $f_{w}=2$ for the position of the gel point is
tested by computing the size of the largest cluster of concatenated
rings in the samples. The result is shown in Figure \ref{fig:The-weight-fraction}.
According to the simulation data of networks close to the gel point
(see also Figure \ref{fig:The-weight-fraction}), the gel point is
located at a weight average functionality $f_{w,c}=2.12\pm0.03$.

To estimate the the weight fractions of active material and gel, we
use the approach of Miller and Macosko \cite{key-22,key-29} and apply
it to a Poisson distributed set of functionalities with given average
$f_{n}$. Let $P_{out}$ denote the probability of finding a finite
chain when ``looking out'' along one of the $f$ connections (concatenations)
of a ring. The functionality of the connected ring is selected randomly
according to the weight fraction of connections to rings with $f$
connections, 
\begin{equation}
a_{f}=\frac{P(f,f_{n})f}{\sum_{f}P(f,f_{n})f}=\frac{f_{n}^{f-1}}{(f-1)!}\mbox{e}^{-f_{n}}=P(f-1,f_{n}).\label{eq:a_f}
\end{equation}
Above we made use of $\sum_{f}P(f,f_{n})f=f_{n}$. Note that the situation
we analyze here is equivalent to full conversion $p=1$ in Refs. \cite{key-22,key-29}
and thus, looking ``out'' or ``into'' a molecule from a given
connection is equivalent: 
\begin{equation}
P_{out}=P_{in}=\sum_{f}a_{f}P_{out}^{f-1}.\label{eq:P_out}
\end{equation}
Thus, we need to solve numerically 
\begin{equation}
\sum_{f}a_{f}P_{out}^{f-2}-1=0\label{eq:sum_f}
\end{equation}
for $P_{out}$ as function of $f_{n}$, which is the only variable
here. We seek a solution of this equation within the interval $[0,1[$,
if existing; otherwise $P_{out}=1$. The weight fraction of sol, $w_{sol}$,
and gel, $w_{gel}$, are computed by inserting this particular solution
for $P_{out}$ into the following equations: 
\begin{equation}
w_{sol}(f_{n})=\sum_{f}P(f,f_{n})P_{out}^{f}\label{eq:w_sol}
\end{equation}
\begin{equation}
w_{gel}(f_{n})=1-w_{sol}\label{eq:w_gel}
\end{equation}
The weight fraction of the active material, $w_{act}$, is the weight
fraction of all rings in gel with at least two independent connections
to the gel 
\begin{equation}
w_{act}(f_{n})=w_{gel}(f_{n})-\sum_{f}P(f,f_{n})fP_{out}^{f-1}(1-P_{out}).\label{eq:w_act}
\end{equation}

The above predictions are compared with simulation data in Figure
(\ref{fig:The-weight-fraction}). We observe good agreement with our
mean field estimate for both weight fraction of gel and active material.
Note that in contrast to conventional gelation the functionality of
the molecules is not determined by chemistry; instead it results from
the interpenetration and concatenation of overlapping rings. The external
parameters $N$ and $\phi$ enter directly via $f_{n}$ in the sol-gel
transition and not in corrections for intra-molecular reactions as
typical for chemically linked gels. Corrections to gelation (e.g.
shift of gel point) are also expected to be universal here, since
such corrections can only arise from multiple links between overlapping
rings, that depend in the same manner from overlap between the molecules
as concatenation. Note that the above analysis allows to determine
$f_{w}$ and thus, $\gamma$ from experimental data in the region
just above the gel point by measuring the weight fraction of solubles.
Finally, we would like to point out that OGs with $f_{n}\gtrsim8$
can be considered essentially as defect free model systems, since
$w_{gel}>0.999$ and $w_{act}>0.99$. This ideal case of interpenetrating
solutions is used in the following two sections below as reference
to analyze DNA-Origami and progressive construction.

\section{\label{sec:DNA-Origami}DNA-Origami}

To analyze the competition between the growth of linear chains and
ring closure, we equilibrated melts of linear chains with rather small
degrees of polymerization $N=16,$ 32, and 64 in order to achieve
significant amounts of ring molecules upon randomly linking all chain
ends. The total number of monomers in each sample was $2^{20}$. The
linear chains were randomly assigned to be part of one of the $B=1$,
2, 4, 8, 16, 32, or 64 batches of chains. Polymer volume fraction
was kept constant at $\phi=0.5$ and all samples were simulated on
a lattice of $256^{3}$ lattice sites. A permanent bond is introduced
whenever two previously unreacted chain ends of the same batch hit
each other during the course of their motion at the smallest possible
separation on the lattice. The extent of reaction at the end of the
simulations was typically above 99\% of the maximum possible extent
of reaction. However, all samples were analyzed at conversion of $p=0.9568$,
which is the smallest maximum conversion that was achieved within
all samples, in order to eliminate systematic effects caused by a
variation of $p$. The number and weight fractions of linear chains
and cyclic polymers of different size were analyzed at the end of
the simulations.

\begin{figure}
\begin{centering}
\includegraphics[angle=270,width=1\columnwidth]{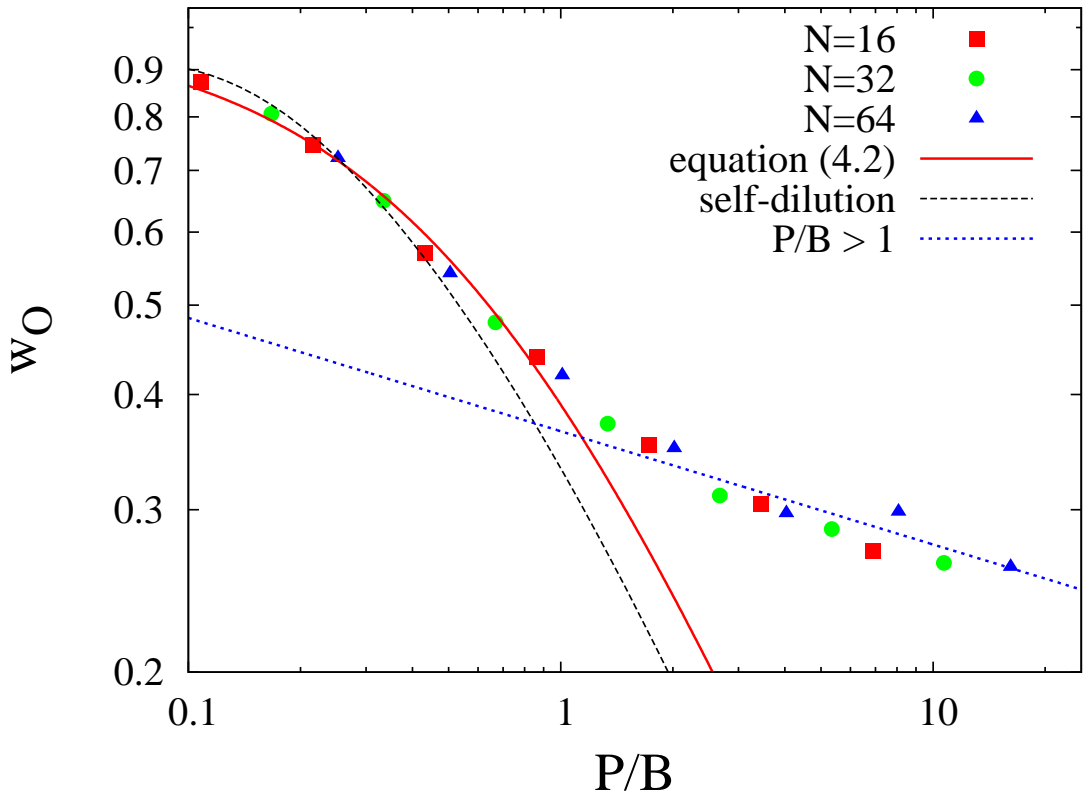}
\par\end{centering}

\caption{\label{fig:Weight-fraction-of}Weight fraction of cyclic polymers
as function of the number of batches $B$ rescaled to overlap concentration.
The ``self-dilution'' estimate was computed for $p=0.9568$ and
fit to the data for with $w_{O}>1/2$, while the effect of $p<1$
was ignored for equation (\ref{eq:Pickett}). The dotted line is a
power law approximation for the tail at large $P/B$ as discussed
in the text.}
\end{figure}

Let us assume that at the beginning of the reaction, all chains are
mono-disperse with a degree of polymerization $N\gg1$ such that ring
formation is not restricted within individual chains. The overlap
number 
\begin{equation}
P\approx\phi R_{g}^{3}/N\label{eq:overlap number}
\end{equation}
describes the number of chains in the pervaded volume of a polymer
at the beginning of the reactions. Splitting the chains into $B$
batches of chains that react selectively within the same batch reduces
the overlap number to $\approx\phi R_{g}^{3}/(BN)=P/B$. Pickett \cite{key-5}
expects that the weight fraction of rings among all polymer, $w_{O}$,
can be described by a function of form 
\begin{equation}
w_{O}\approx\frac{1}{1+zP/B}\text{,}\label{eq:Pickett}
\end{equation}
whereby $z$ is a numerical constant close to two reflecting the fact
that only one opposite chain end is in the vicinity for ring formation,
while each overlapping linear chain contributes two reactive groups.
Our simulation data in the ``dilute'' regime $B>P$ is well described
by this prediction with $z\approx1.60\pm0.05$ as adjustable parameter,
see Figure \ref{fig:Weight-fraction-of}. However, the available data
at $P/B>1$ seems to be a function of $P/B$, such that 
\begin{equation}
w_{O}\approx y\left(P/B\right)^{\alpha}\label{eq:w_O}
\end{equation}
with $\alpha\approx-0.12\pm0.03$ and $y\approx0.36\pm0.02$.

We searched literature for explanations of the regime $P/B>1$, but
neither a numerical test \cite{key-28} of the compact exploration
of space of the reactive chain ends \cite{key-27} (resulting in a
much larger $\alpha$) nor an explicit computation of the self-dilution
effect \cite{key-26,key-25} by using the mean field rate equations
of the Appendix lead to convincing results (see Figure \ref{fig:Weight-fraction-of}).

\begin{figure}
\includegraphics[angle=270,width=1\columnwidth]{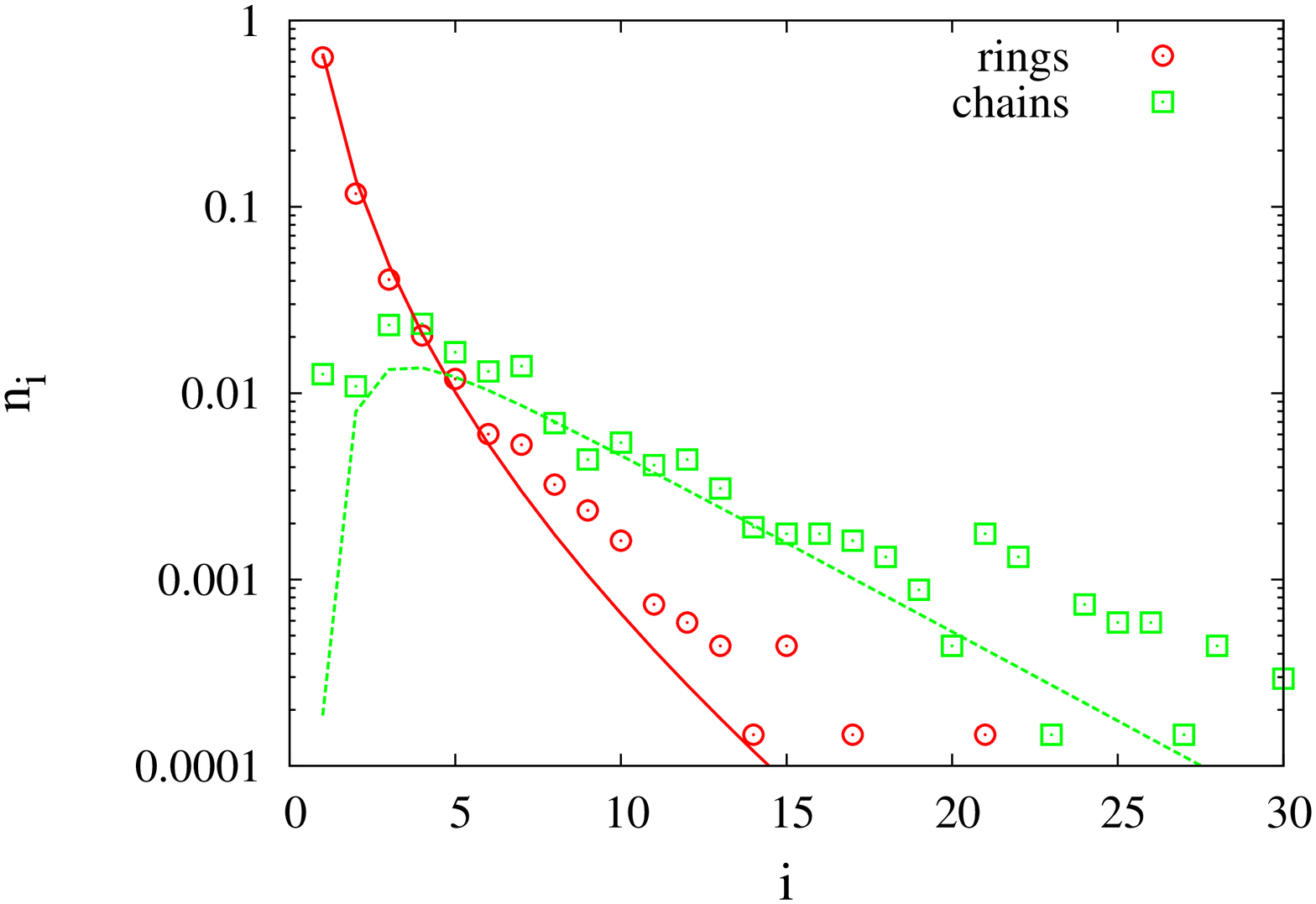}

\caption{\label{fig:P/B<1}Typical number fraction distribution $n_{i}$ of
linear chains and cyclic polymers made of $i$ precursor chains at
$P/B<1$. Data taken at $p=0.9568$ of a sample with $N=64$ and $B=32$.
Lines are computed using the differential equations described in the
Appendix. }
\end{figure}

\begin{figure}
\includegraphics[angle=270,width=1\columnwidth]{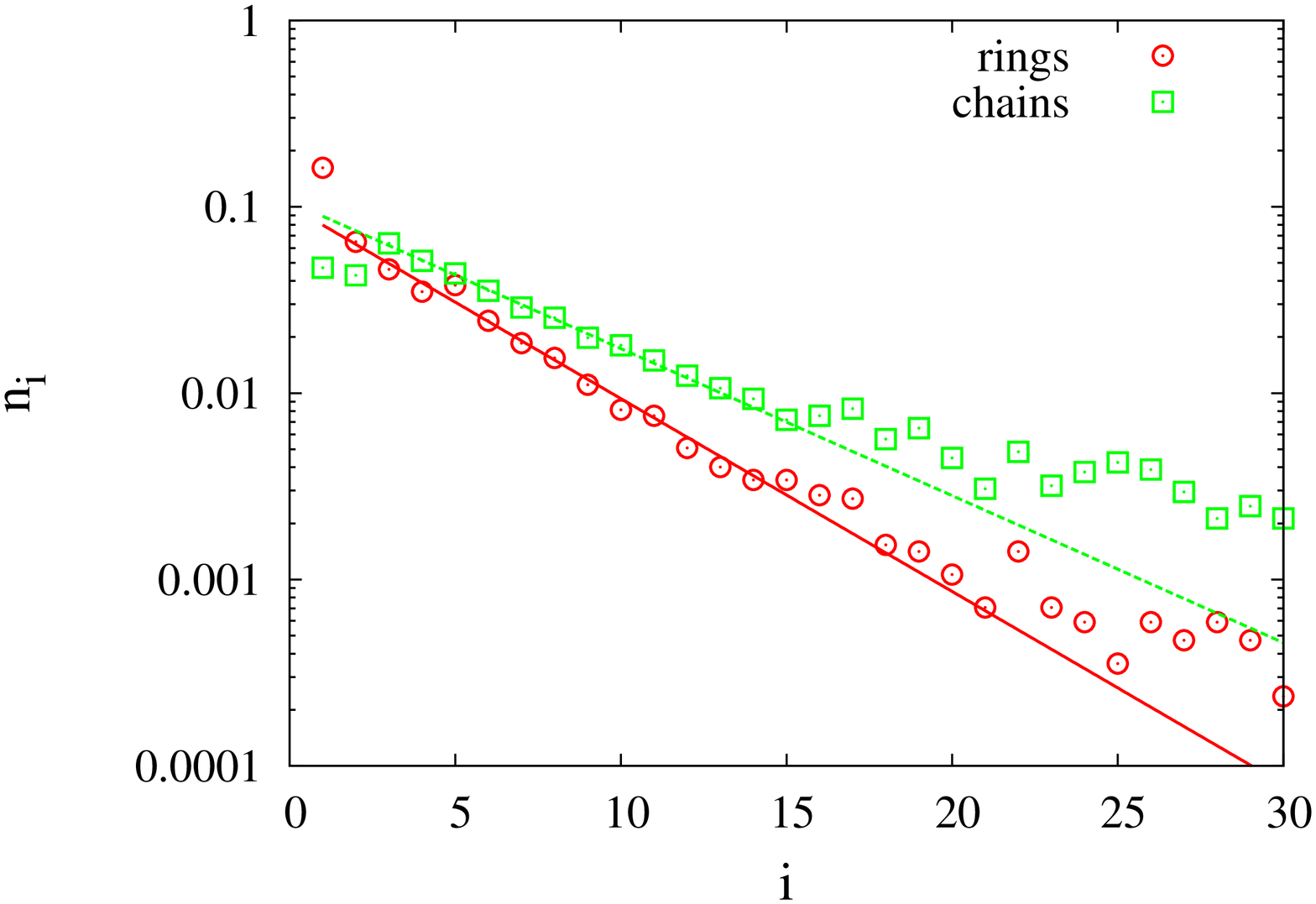}

\caption{\label{fig:P/B>1}Typical number fraction distribution $n_{i}$ of
linear chains and cyclic polymers at $P/B>1$. Data taken at $p=0.9568$
of a sample with $N=16$ and $B=1$. Lines are fits to functions $ap^{i-1}(1-p)$
with variables $a$ and $p$. }
\end{figure}

In addition to the above trends for the total weight fraction of cyclic
polymers, $w_{O}$, we observe a qualitative change in the number
fraction distributions $n_{i}$ of linear chains and rings made of
$i$ precursor chains for the two regimes $P/B>1$ and $P/B<1$ as
shown in Figure \ref{fig:P/B<1} and \ref{fig:P/B>1}. The data below
overlap, $P/B<1$ can be approximated by the set of differential equations
given in the appendix. The effect of self-dilution is visible here
by the depletion of shortest linear chains as compared to a most probable
distribution (linear decay on semi-log plot). The decay of the number
fraction of rings at small $i$ can be approximated alternatively
by a power law with an exponent near $-3/2$ with an exponential cut-off
similar to the distribution of the linear chains. In contrast to this,
both distributions apparently become two most probable distributions
with different effective conversions $p$ in the regime $P/B>1$ in
a first order approximation, see Figure \ref{fig:P/B>1}. Here, the
effective $p$ to describe the distributions of the rings is always
smaller as the one of the linear chains, which is close to equation
(\ref{eq:p-1}) of the Appendix. Note that in both cases an efficient
formation of concatenated rings is only possible, if the degree of
polymerization of the precursor linear chains is already above the
cut-off for non-concatenation.

Let us now discuss the gel point condition and network structure for
DNA origami in both limits $P/B<1$ and $P/B>1$ based upon the available
data. For simplicity, we assume, that the precursor chains have a
higher degree of polymerization as the cut-off degree of polymerization
for concatenation $N>N_{OO}$. This allows us to use the weight fraction
of rings $w_{O}$ as correction for the average number of concatenations
and to drop the non-concatenation correction in equation (\ref{eq:fn}).

In the limit $P/B<1$ , the number fraction distribution of rings
is dominated by the strong decay at small $i$ such that we can keep
in first approximation $f_{w}\approx f_{n}+1$, which simplifies the
gel point condition to 
\begin{equation}
\gamma\phi^{\nu(3\nu-1)}Nw_{O}\approx f_{n}\approx1.\label{eq:gel}
\end{equation}
Under these conditions, all results of the previous section can be
kept as first order approximation, after the average number of concatenations
is corrected by a factor $w_{O}$ as in the above equation. The number
of batches $B$ necessary to pass through the gel point is here 
\begin{equation}
B\approx\frac{zN^{1/2}}{\gamma\phi^{\nu(3\nu-1)}N-1}\label{eq:B}
\end{equation}
because of $P\approx N^{1/2}$. Since $P/B<1$, there is $B\gtrsim N^{1/2}$,
which yields that the above criterion holds for $N\phi^{\nu(3\nu-1)}\lesssim(z+1)/\gamma$.
All samples with larger $N\phi^{\nu(3\nu-1)}$ are gelled, if $B>P$.

Note that the above two equations can be used to map the DNA Origami
in the regime $P/B<1$ in first approximation back onto the ideal
OG case. For instance, considering our results for the ideal OGs we
conclude that almost defect free model networks (after removal of
linear chains) with a weight fraction of $w_{net}\approx w_{O}$ are
obtained for $P/B<1$, if $f_{n}\gtrsim8$.

In the limit $P/B>1$, the number fraction distribution of rings is
roughly given by a most probable weight distribution, see Figure \ref{fig:P/B>1}.
Because of the narrowly distributed functionalities for a given $N$,
equation (\ref{eq:poisson}), the relation between number and weight
average functionality is dominated by the broader distribution of
the molecular weights. Thus, $f_{w}\approx2f_{n}$. Thus, the gel
point condition is the same as in equation (\ref{eq:gel}) by coincidence.
The weak dependence of $w_{O}$ on $P/B$ for $P/B>1$ shows that
a variation of the number of batches for reaction is not very effective
unless one reaches the opposite regime $P/B<1$. On the other hand,
since the average functionality of the rings grows $\propto\gamma\phi^{\nu(3\nu-1)}N^{1-\alpha/2}$
for constant $B$, the probably most efficient strategy in this limit
$P/B>1$ is simply to choose a precursor degree of polymerization
$N\gg N_{OO}$ such that $f_{n}>1$ despite the low $w_{O}$. However,
this yields low network weight fractions $w_{net}<w_{O}\ll1$ and
a high fraction of long linear chains $1-w_{O}$ needs to be extracted.
As a result, one might obtain a largely de-swollen Olympic network.

We have to comment here, that the observed power law for $w_{O}$
as function of $B/P$ must not necessarily extend to ratios of $P/B$
much larger than the ones obtained in our study - as it would be necessary
to achieve gelation. The worst case scenario is here $w_{O}\propto(P/B)^{-1}$,
which still leads at a constant $B$ and $\phi$ to a growing functionality
of the rings, $f_{n}\propto N^{1/2}$, if $N$ remains below an $N^{*}$
above which any pair of overlapping rings is concatenated (see Ref.
\cite{key-11} for details). For $N>N^{*}$ on the other hand, $f_{n}$
becomes constant in this worst case scenario and increasing $N$ beyond
$N^{*}$ will not cause the formation of an OG. However, for the worst
case scenario $w_{O}\propto(P/B)^{-1}$, a variation of $B$ is very
effective, since $f_{n}\propto B$ for both regimes $N<N^{*}$ and
$N>N^{*}$.

\section{\label{sec:Progressive-construction}Progressive construction}

The results of the two previous sections can be applied to discuss
the progressive construction methods for OGs. The original approach
of Ref \cite{key-4} is devoted to a mixture of long and short polymers
that are both below the entanglement length in order to avoid complications
by the compression of the long cyclic polymers. This requires that
the concatenation degree of polymerization $N_{OO}$ is clearly below
the entanglement degree of polymerization $N_{e}$. In our previous
work \cite{key-11,key-13}, we rather find the opposite: $N_{OO}\gtrsim N_{e}$.
Even ideal mono-disperse samples (see section \ref{sec:Ideal-Olympic-gels})
do not form OGs for $N<N_{e}\lesssim N_{OO}$. Therefore, we develop
an alternative approach for progressive construction at $N>N_{OO}$
using the same strategy as proposed in Ref. \cite{key-4}: we use
short linear strands to link long non-concatenated rings that were
concentrated above overlap concentration.

Let us use $L$ and $S$ to denote the degrees of polymerization of
the long and short molecules respectively. Since the Kuhn molecular
weight of the simulation model is close to one monomer, we drop all
coefficients related to converting monomers to Kuhn monomers. Next,
$w_{L}$ and $w_{S}$ denote the weight fractions of long and short
rings among all polymer, while $w_{lin}$ is the weight fraction of
linear chains that is obtained as by-product of the linking reactions.
Thus, $w_{L}+w_{S}+w_{lin}=1$. Finally, $\phi$ is the weight fraction
of polymers, $b$ the root mean square length of a segment and the
monomeric volume $\mbox{v}_{0}$ is $\approx b^{3}$.

First, let us consider the case $w_{S}\ll w_{L}\approx1$ so that
conformational changes upon the addition of short polymers can be
ignored in first approximation. Furthermore, we add short chains well
below their overlap volume fraction such that $w_{S}>w_{lin}$ and
the cyclic chains produced by the $S$-mers consist predominantly
of only one $S$-mer. Since the $S$-mers are dilute, we assume that
$S$-mers entrap in first approximation only $L$-mers \cite{key-24}.
Thus, 
\begin{equation}
f_{n,S}\approx\gamma\phi^{\nu/(3\text{\ensuremath{\nu}-1})}S(w_{L}+w_{S})\approx\gamma\phi^{\nu/(3\text{\ensuremath{\nu}-1})}S.\label{eq:fns}
\end{equation}
These concatenations are distributed among all $L$-mers, for which
we obtain a concatenation density that is reduced by a factor of $w_{S}/w_{L}$.
This lower concatenation density determines the average number of
concatenations of the long chains 
\begin{equation}
f_{n,L}\approx\frac{w_{S}}{w_{L}}\gamma\phi^{\nu/(3\text{\ensuremath{\nu}-1})}L.\label{eq:fnl}
\end{equation}
In both cases, we expect that the number of concatenations are Poisson
distributed \cite{key-33}. For concatenations only between long and
short rings we can map the problem to co-polymerizations of molecules
with distributed functionalities that are discussed in Ref. \cite{key-29}.
Then, the gel point condition can be written as 
\begin{equation}
(f_{w,S}-1)(f_{w,L}-1)=f_{n,S}f_{n,L}\approx\frac{w_{S}}{w_{L}}\frac{L}{S}f_{n,S}^{2}=1.\label{eq:gel point}
\end{equation}

Thus, $w_{L}\gg w_{S}$ requires that $Lf_{n,S}^{2}/S\gg1$, which
is best obtained for $f_{n,S}\gtrsim1$ in order to avoid gigantically
long polymers $L$, since for very small $f_{n,S}$ the exponential
cut-off for concatenation \cite{key-11} would lead to a square exponential
growth of $L$. In contrast to this result, $f<1$ has been suggested
in Ref. \cite{key-4} for the small rings to create OGs. Another interesting
point of equation (\ref{eq:gel point}) is that it allows to determine
$\gamma$ experimentally, if the polydispersity of the rings made
by $S$-mers is low, since then, all parameters of this equation except
of $f_{n,S}^{2}$ and thus, $\gamma$ are known.

Similar to section \ref{sec:Ideal-Olympic-gels}, the final goal is
to achieve OGs with a well developed network structure such that essentially
all $L$-mers are active. Since linking of long chains should happen
by a volume fraction of short chains below their overlap volume fraction
(note that $S$-mers are larger than blob size at $f_{n,S}>1$), we
require 
\begin{equation}
w_{S}\phi<\phi_{S}^{*}\approx\frac{b^{3}S}{\left(bS^{1/2}\phi^{-(\nu-1/2)/(3\nu-1)}\right)^{3}},\label{eq:wsphi}
\end{equation}
which is 
\begin{equation}
w_{S}\lesssim S^{-1/2}\phi^{-(6\nu-5/2)/(3\nu-1)}\label{eq:ws}
\end{equation}
as upper limit for $w_{S}$. Let us use this upper limit as bound
for the maximum weight fraction to be added to obtain a ``well developed''
network. Using the results of the ideal OGs of section \ref{sec:Ideal-Olympic-gels}
as reference, let us adopt the following criteria $f_{n,L}f_{n,S}\ge8$
and $f_{n,S}\ge1$ to ensure that essentially all large rings are
incorporated into the gel. Thus, we require 
\begin{equation}
\frac{w_{S}}{w_{L}}\frac{L}{S}f_{n,S}^{2}\ge8\label{eq:wswl}
\end{equation}
at the end of the reactions. A weight fraction $w_{S}$ that fulfills
equation (\ref{eq:ws}) and (\ref{eq:wswl}) can be found, if 
\begin{equation}
L\ge8w_{L}S^{-1/2}\gamma^{-2}\phi^{-(5/2-4\nu)/(3\nu-1)},\label{eq:L}
\end{equation}
which requires an enormous $L\gtrsim900$ for our simulations to start
with, because here $\gamma^{-1}\approx30$ and $S\ge50$ for $f_{n,S}\ge1$
(at $\phi=0.5$). This result shows that the construction of well
developed OGs is possible within a single concatenation step, given
that sufficiently long chains $L$ are available to be linked.

More details about network structure can be obtained numerically using
the approach of Miller and Macosko \cite{key-22,key-29} applied to
co-polymerization similar to our discussion in section \ref{sec:Ideal-Olympic-gels}.
One interesting point is here, that the weight fraction of sol could
be used for an alternative determination of $f_{n,S}$ and thus $\gamma$
if $f_{n,S}$ is sufficiently large such that the exponential cut-off
for $f_{n,S}$ could be ignored. This is because $f_{n,S}$ is Poisson
distributed and a fraction of $e^{-f_{n,S}}$ of the short chains
will not entrap any polymer. Thus, well beyond the gel point, $f_{n,S}f_{n,L}\gg1$,
the weight fraction of sol will be dominated by non-concatenated short
chains and linear strands, $w_{sol}\approx w_{lin}+w_{S}e^{-f_{n,S}}$,
which may be used to estimate $f_{n,S}$, if $w_{lin}$ is sufficiently
small or linear chains can be separated from cyclic polymers.

A combination of progressive construction and DNA-Origami would allow
to increase the lower boundary for the weight fraction of the short
polymers, $w_{S}$, in equation (\ref{eq:ws}) by a factor equal to
the number of batches, \textbf{$B$}, such that the required chain
length $L$ is reduced to $L/B$. However, a too large $B$ will break
the assumption \cite{key-23} $w_{S}\ll w_{L}$ which is for $B=1$
essentially always satisfied because of the low value of $\gamma$.
Instead, if a large $B$ is available, it could be used to minimize
$w_{lin}$ instead in order to remove effects of poly-disperse rings
made of $S$-mers and to reduce the weight fraction of sol.

On the other hand, the combination of DNA-Origami and progressive
construction can be used to reduce the weight fraction of sol for
a given possible maximum number of batches $B$ as compared to pure
DNA-Origami. As trade off one obtains a sample that consists of partially
compressed long chains concatenated with short chains that are at
equilibrium at cross-linking conditions, which is the general complication
of creating OGs by progressive construction. Indeed, for progressively
constructed gels one has to expect an even more unusual swelling behavior
as found recently for ideal OGs \cite{key-3}, since beyond dis-interpenetration
of rings, the long rings gain extra conformations by reducing their
compression.

\section{\label{sec:Summary}Summary}

In the present paper we have discussed analytically and numerically
the formation and structure of Olympic gels (OGs) and tested our predictions
by simulation data. We focused on three different model cases: ideal
OGs, DNA-Origami, and progressive construction. For ideal OGs we demonstrate
that the distribution of the number of concatenations is well approximated
by a Poisson distribution. This Poisson distribution can be used to
predict numerically structural features of the gels, such as the gel
point, the weight fraction of gel, or the weight fraction of elastically
active rings. These predictions are well supported by simulation data.
While the construction of ideal OGs was not considered explicitly
in previous work, we found for the other two cases clear differences
to previous works \cite{key-4,key-5}.

Below overlap concentration of the linear chains used for DNA-Origami,
the model of Pickett \cite{key-5} is well suited to describe the
weight fraction of rings. Above overlap, we find a much smaller decay
of the weight fraction of rings as function of the overlap number
$P$ and the number of batches $B$ with selectively binding ends
that is $\propto\left(P/B\right)^{\alpha}$ with an $\alpha\approx-0.12\pm0.03$.
A more detailed mean field analysis taking into account the competition
between ring formation and growth of linear chains does not lead to
an improved description of the simulation data. However, following
the results of the mean field model one can conclude that by far the
most efficient way to achieve a large weight fraction of rings is
to choose a low overlap $P/B$ of the linear chains at the onset of
the reaction. Since the weight fraction of rings is dominated by the
smallest rings, possible steric cut-offs for ring formation and concatenation
seriously affect the formation of rings that can concatenate.

Progressive construction can be used indeed, to obtain OGs. However,
the conditions necessary for gelation are rather opposite to the ones
proposed originally by \cite{key-4}: instead of using very short
strands with a number average number of concatenations $f_{n}<1$,
we suggest to use $f_{n}>1$, since otherwise, the required minimum
degree of polymerization of the long rings, $L$, will grow square
exponential when further reducing $f_{n}$. The Poisson distribution
for the number of concatenations allows again to derive a rather simple
condition for gelation. However, the low numerical constant $\gamma$
for concatenation leads to still very large $L$ to obtain well developed
OGs, if the weight fraction of the short polymers is low in order
to not disturb the conformations of the long chains.

In general, DNA-Origami can be combined with progressive construction.
This allows to either use a smaller number of batches as compared
to DNA-Origami, or a smaller $L$ could be used in progressive construction,
which may simplify the construction of OGs.

\section{Acknowledgement}

The authors acknowledge a generous grant of computing time at the
ZIH Dresden for the project BiBPoDiA. Financial support was given
by the DFG grants LA 2735/2-1 and SO 277/7-1.

\section{Appendix: }

\subsection{\label{sec:Appendix-A:-Mean}Mean field approximation of ring-chain
competition}

The competition between the growth of linear chains and ring formation
can be analyzed numerically in a mean-field framework by an adequate
set of differential equations. We consider first the linear condensation
of short $N$-mer chains (monomers ``A'') without ring formation.
Let $n_{i}(p)$ denote the number fraction of $iN$-mers at conversion
$p$. At the absence of ring formation, one obtains for irreversible
linear condensation reactions of monomers A 
\begin{equation}
(\mbox{A})_{i}+(\mbox{A})_{j}\rightarrow(\mbox{A})_{i+j}\label{eq:An}
\end{equation}
a most probable weight distribution with the well known polymer number
fraction distribution 
\begin{equation}
n_{i}(p)=p^{i-1}(1-p)\label{eq:nN}
\end{equation}
and an average degree of polymerization that is here 
\begin{equation}
N_{\mbox{n}}=N\frac{1}{1-p}.\label{eq:Nn}
\end{equation}

A numerical solution of this problem can be computed using the following
set of differential equations: The number fraction $di^{-}(p)$ of
$i$-mers that react within a integration interval $\mbox{d}p$ is
\begin{equation}
\mbox{d}i^{-}(p)=n_{i}(p)\mbox{d}p,\label{eq:dN-}
\end{equation}
while the fraction $di^{+}(p)$ of $i$-mers formed during the integration
interval is given by 
\begin{equation}
\mbox{d}i^{+}(p)=\sum_{k=1}^{i-1}n_{k}(p)n_{i-k}(p)\mbox{d}p.\label{eq:dN+}
\end{equation}
Since the distribution $n_{i}(p)$ is normalized to one, the change
in conversion equals $\mbox{d}p$. With initial conditions $n_{1}(0)=1$
and $n_{i}=0$ for $i>1$ we obtain numerically equation (\ref{eq:Nn}),
by computing 
\begin{equation}
n_{i}(p+\mbox{d}p)=n_{i}(p)+\mbox{d}i^{+}(p)-\mbox{d}i^{-}(p)\label{eq:ni}
\end{equation}
in infinitesimal integration intervals $\mbox{d}p$. The analytical
solution, equation (\ref{eq:Nn}), can be used to check the accuracy
of the numerical solution.

Ring formation is implemented using the same assumptions (Gaussian
statistics for all $iN$-mers, mean field approximation) as in section
\ref{sec:DNA-Origami}. The average concentration of the reactive
groups $c(p)$ is proportional to the initial concentration $c_{0}$
of reactive groups and decays with conversion 
\begin{equation}
c(p)=c_{0}(1-p).\label{eq:cp}
\end{equation}
Gaussian statistics for all precursor chains and combined linear chains
of $i$ sections with $N$ monomers leads to concentrations 
\begin{equation}
c_{i}\approx c_{1}i^{-3/2}\label{eq:cN}
\end{equation}
of the first end of an $iN$-mer near its second end. Here $c_{1}\approx1/(2P)$
is the corresponding concentration for one chain of $N$ monomers.
Assuming equal reactivity, all reaction rates are proportional to
the concentrations of the reactive species only. Thus, we can equate
for the rate to form a ring polymer of $iN$ monomers, $dC_{i}^{+}$,
that 
\begin{equation}
\frac{\mbox{d}r_{i}^{+}(p)}{\mbox{d}i^{-}(p)}=\frac{c_{i}}{c_{0}(1-p)}.\label{eq:dC}
\end{equation}
These additional reactions that convert linear chains of $iN$ monomers
into cycles disturb the most probable distribution of the linear species.
Nevertheless, the total number fraction of all species is still normalized,
\begin{equation}
\sum_{i}n_{i}(p)+\sum_{i}r_{i}(p)=1,\label{eq:normation}
\end{equation}
whereby only the number fraction of linear chains, 
\begin{equation}
n_{lin}(p)=\sum_{i}n_{i}(p),\label{eq:linear fraction}
\end{equation}
is available for further reactions. Note that in the above equations
$d_{i}^{+}(p)$ is $\propto n_{lin}^{2}(p),$ while $d_{i}^{-}(p)$
and $dr_{i}^{+}(p)$ are proportional to $n_{lin}(p)$. Therefore,
we have to modify equation (\ref{eq:dN+}) to 
\begin{equation}
\mbox{d}i^{+}(p)=\sum_{k=1}^{i-1}n_{k}(p)n_{i-k}(p)\mbox{d}p/n_{lin}(p),\label{eq:di}
\end{equation}
if there is ring formation in order to maintain normalization of the
distributions. The number fractions of rings and linear chains are
obtained by computing 
\begin{equation}
r_{i}(p+\mbox{d}p')=r_{i}(p)+\mbox{d}r_{i}^{+}(p)\label{eq:ri}
\end{equation}
\begin{equation}
n_{i}(p+\mbox{d}p')=n_{i}(p)+\mbox{d}i^{+}(p)-\mbox{d}i^{-}(p)-\mbox{d}r_{i}^{+}(p).\label{eq:ni-1}
\end{equation}
Note that the effective infinitesimal conversion $\mbox{d}p'$ is
affected by ring forming reactions leading to 
\begin{equation}
dp'=\sum_{i}\left(\mbox{d}i^{-}(p)+\mbox{d}r_{i}^{+}(p)\right).\label{eq:dp'}
\end{equation}
Ring formation also modifies the conversion of the linear species,
since $p\equiv1$ for all rings. Let $w_{O}(p)$ denote the weight
fraction of rings at conversion $p$. The conversion among the linear
chains, $p_{lin}$, is thus given by

\begin{equation}
p_{lin}=\frac{p-w_{O}(p)}{1-w_{O}(p)}.\label{eq:p-1}
\end{equation}
This relation can be used to compute $p$ directly from the average
degree of polymerization of the linear chains, $N_{\mbox{n}}$, by
inserting $p_{lin}$ instead of $p$ into equation (\ref{eq:Nn})
for the linear species. 

\begin{thebibliography}{10}
\bibitem{key-1}P. G. de Gennes, ``Scaling Concepts In Polymer Physics'',
Cornell University Press, New York, NY, USA (1991).

\bibitem{key-2}T. A. Vilgis and M. Otto, \emph{Phys. Rev. E} \textbf{56},
R1314 (1997).

\bibitem{key-3}M. Lang, J. Fischer, M. Werner, and J.-U. Sommer,
\emph{Phys. Rev. Lett.} \textbf{112}, 238001 (2014).

\bibitem{key-4}E. Raphael, C. Gay, and P. G. de Gennes, \emph{J.
Stat. Phys. }\textbf{89}, 111 (1997).

\bibitem{key-5}G. T. Pickett, \emph{Europhys. Lett.} \textbf{76},
616 (2006).

\bibitem{key-6}I. Carmesin and K. Kremer, \emph{Macromolecules} \textbf{21},
2819 (1988).

\bibitem{key-7}K. Binder and H. Deutsch,\emph{ J. Chem. Phys.} \textbf{94},
2294 (1991).

\bibitem{key-8}J. P. Wittmer, P. Beckrich, H. Meyer, A. Cavallo,
A. Johner, and J. Baschnagel,\emph{ Phys. Rev. E} \textbf{76}, 11803
(2007).

\bibitem{key-9}M. Lang, M. Rubinstein, and J.-U. Sommer, \emph{ACS
Macro Letters}\textbf{ 4}, 177 (2015). 

\bibitem{key-10}W. Paul, K. Binder, D. W. Heermann, and K. Kremer,\emph{
J. Phys. II France}\textbf{\emph{ }}\textbf{1}, 37 (1991). 

\bibitem{key-11}M. Lang, J. Fischer, and J.-U. Sommer, \emph{Macromolecules}
\textbf{45}, 7642 (2012). 

\bibitem{key-12}M. Lang and J.-U. Sommer, \emph{Phys. Rev. Lett.
}\textbf{104}, 177801 (2010). 

\bibitem{key-13}M. Lang, \emph{Macromolecules} \textbf{46}, 9782
(2013). 

\bibitem{key-14}M. Lang, W. Michalke, and S. Kreitmeier,\emph{ J.
Comp. Phys.} \textbf{185}, 549 (2003). 

\bibitem{key-15}G. Gouesbet, S. Meunier-Guttin-Cluzel, and C. Letellier,
\emph{Applied Mathematics and Computation} \textbf{105}, 271 (1999).

\bibitem{key-16}P. Freyd, D. Yetter, J. Hoste, W. Lickorish, K. Millet,
and A. Ocneanu, \emph{Bull. Am. Soc. }\textbf{12}, 239 (1985). 

\bibitem{key-17}W. Michalke, M. Lang, S. Kreitmeier, and D. Göritz,
\emph{Phys. Rev. E} \textbf{64}, 012801 (2001). 

\bibitem{key-18}M. Tanaka, K. Iwata, and N. Kuzuu, Comp. \emph{Theo.
Pol. Sci.} \textbf{10}, 299 (2000). 

\bibitem{key-19}P. J. Flory, ``Principles Of Polymer Chemistry'',
Cornell University Press, Cornell, NY, USA (1953). 

\bibitem{key-20}M. Lang, \emph{Macromolecules} \textbf{46}, 1158
(2013).

\bibitem{key-32}The Poisson distribution is the mathematical approximation
of independent random events which occur at low average rate. By using
this approximation, we thus, implicitly ignore additional conditions
to the underlying random process as may result from packing conditions
and neglect possible correlations among concatenations: For instance,
each concatenation blocks a certain volume near the minimal surface
for other cyclic molecules and thus, a weak anti-correlation (repulsion)
among concatenating molecules is to be expected, which should lead
to a narrowing of the distributions. Since the deviations between
data and Poisson distribution in Figure \ref{fig:Distribution-of-the}
are rather small, these additional effects are either ignorable or
compensated by the neglect of the minimal area distribution of the
rings mentioned previously.

\bibitem{key-21}L. H. Peebles, ``Molecular Weight Distributions
in Polymers'', Wiley, New York, USA (1971). 

\bibitem{key-22}D. Miller and C. Macosko, \emph{Macromolecules} \textbf{9},
206 (1976). 

\bibitem{key-29}C. Macosko and D. Miller, \emph{Macromolecules} \textbf{9},
199 (1976). 

\bibitem{key-28}Y.-K. Leung and B. Eichinger, \emph{J. Chem. Phys.
}\textbf{80}, 3885 (1984). 

\bibitem{key-27}P. de Gennes, \emph{J. Chem. Phys. }\textbf{76},
3316 (1982). 

\bibitem{key-26}H. R. Kricheldorf, S. Böhme, and G. Schwarz, \emph{Macromolecules}
\textbf{34}, 8879 (2001). 

\bibitem{key-25}H. R. Kricheldorf and G. Schwarz, \emph{Macromol.
Rapid Commun. }\textbf{24}, 359 (2003). 

\bibitem{key-24}Actually, there is an entropic bias towards penetration
of long chains, if the long chains are compressed. 

\bibitem{key-33}Here, concatenations of $L$ rings are caused only
by $S$ rings, which are dilute in the melt of $L$ rings. Thus, the
positions of the $S$ rings are not correlated and the rate of concatenation
for line elements of the $L$ rings are low. Therefore, the $L$ rings
must show a Poisson distribution of concatenations. The concatenation
distribution of $S$ rings, on the other hand, can be mapped onto
the monodisperse ideal Olympic gel case, since the number of strands
of large $L$ rings that overlap with the $S$ ring is of order $S^{1/2}$.
However, packing constraints might be somewhat relaxed here, which
could lead to a slightly broadened distribution in the number of concatenations
as compared to the monodisperse case.

\bibitem{key-23}And thus, swell the long chains and cause a signicant
fractions of S-mer concatenations by S-mers. This may complicate a
theoretical description of these samples, but it will clearly allow
to obtain well developed OGs at a lower L. 

\end{thebibliography}
\end{document}